\nonstopmode

\documentclass[apl,preprint]{revtex4}
\usepackage{amssymb}
\usepackage{amsmath}
\usepackage{graphicx}
\usepackage{caption}
\begin{document}
\title{Long-range coherent coupling in a quantum dot array}

\author{Floris R. Braakman$^{1\ast}$}
\author{Pierre Barthelemy$^{1}$}
\author{Christian Reichl$^{2}$}
\author{Werner Wegscheider$^{2}$}
\author{Lieven M. K. Vandersypen$^{1\ast}$}
\affiliation{1: Kavli Institute of Nanoscience, TU Delft, 2600 GA Delft, The Netherlands}
\affiliation{2: Solid State Physics Laboratory, ETH Z\"{u}rich, 8093 Z\"{u}rich, Switzerland}
\date{\today}

\begin{abstract}Controlling long-range quantum correlations is central to quantum computation and simulation. In quantum dot arrays, experiments so far rely on nearest-neighbour couplings only, and inducing long-range correlations requires sequential local operations. Here we show that two distant sites can be tunnel coupled directly. The coupling is mediated by virtual occupation of an intermediate site, with a strength that is controlled via the energy detuning of this site. It permits a single charge to oscillate coherently between the outer sites of a triple dot array without passing through the middle, as demonstrated through the observation of Landau-Zener-St\"{u}ckelberg interference. The long-range coupling significantly improves the prospects of fault-tolerant quantum computation using quantum dot arrays and opens up new avenues for performing quantum simulations in nanoscale devices.\end{abstract}

\maketitle

\textbf{Nanofabricated circuits of quantum dot arrays provide an excellent platform for quantum information processing using single charges or spins~\cite{Loss98,Taylor05,Hanson07,Hayashi03}. In such a dot array architecture, the tunnel coupling between neighbouring dots plays an essential role. It governs the motion of charges between the dots, permitting delocalization~\cite{Oosterkamp98} over the dots and coherent oscillations between them~\cite{Hayashi03,Petta04}, and the same tunnel coupling is at the core of exchange-based quantum gates on spin qubits~\cite{Loss98,Petta05,Nowack11}.} 

\textbf{Tunnel coupling falls off exponentially with distance, and all experiments on quantum dot arrays so far rely on nearest-neighbour couplings only. In addition, quantum dot arrays are typically constructed from one-dimensional segments since realizing two-dimensional arrays is challenging. These restrictions severely constrain the range of experiments possible in this system at present. Instead of having to repeatedly swap neighboring qubits down the chain, a long-range coupling would enable quantum gates between distant qubits in one step, thereby giving access to many of the benefits of a two-dimensional lattice. This would strongly reduce the requirements for fault-tolerant quantum computing~\cite{Stephens09,Fowler09} and permit quantum simulation of phenomena that are otherwise inaccessible in this system, for instance involving frustration~\cite{Kim10} or superexchange~\cite{Recher01,Mattis06}.}

The most common approach to realizing an effective long-range coupling is to use a quantum bus, as demonstrated for trapped ions~\cite{Schmidt-Kaler03} and superconducting qubits~\cite{Majer07,Sillanpaa07}. For quantum dots, such a bus has been proposed in the form of optical cavities~\cite{Imamoglu99} and microwave stripline resonators~\cite{Taylor06,Burkard06,Trif08}. For the latter, the first steps have been taken recently~\cite{Frey12,Petersson12}. Furthermore, charge transfer through a channel connecting two distant quantum dots has recently been realized using surface acoustic waves that push electrons forward~\cite{Hermelin11,Mcneil11}. In this approach charge coherence is lost but spin coherence is expected to be preserved.

As an alternative for creating long-range coupling of quantum dots, which does not require separate elements such as cavities or channels, a quantum bus has been proposed in the form of the continuum of conduction or valence band states~\cite{Trauzettel07}. Through a second-order process known as cotunneling, virtual occupation of these states can induce an effective coupling between distant quantum dots. Inspired by this scheme, we propose to create such long-range coupling by virtual occupation of discrete states of quantum dots located in between. In this case only discrete levels participate in the cotunneling process. This permits a fully coherent process, in contrast to all existing measurements of cotunneling in quantum dots in which quantum coherence is quickly lost in the reservoirs (see~\cite{Ihn10} for a review).

Here we demonstrate the coherent transfer of single electron charges between the outer sites of a linear array consisting of three quantum dots, in a regime where sequential transport through the middle dot is suppressed energetically. Using real-time charge detection techniques, we study the dependence of the rate with which electrons hop between the outer dots on the detuning of the middle dot levels. We observe a non-monotonous dependence that is characteristic of a coupling mechanism mediated by virtual occupation of the middle dot levels. We also control quantum coherent dynamics between the outer dots in the form of Landau-Zener-St\"{u}ckelberg (LZS) interference, induced by a process we dub photon-assisted cotunneling (PACT).

A scanning electron micrograph of a device identical to the one used is shown in Figure 1a. Gate electrodes fabricated on the surface of a GaAs/AlGaAs heterostructure (see Supplementary Information) are biased with appropriate voltages to selectively deplete regions of the two-dimensional electron gas (2DEG) below and define the linear array of three quantum dots. In the array, only adjacent dots are connected through tunnel barriers. The left and right dots are also tunnel coupled to the left and right reservoirs respectively.  Above the blue-shaded gate a charge sensing dot (SQD) is created, the conductance of which is sensitive to the number of charges on each dot in the array through capacitive coupling. For maximum sensitivity, the SQD is operated on the flank of a Coulomb peak. Furthermore, one of the SQD contacts is connected via a bias-tee to an LC-circuit so that the SQC conductance can be measured both by RF reflectometry~\cite{Reilly07} and in DC (see Supplementary Information). 

\begin{figure}[htb]
\centering
\includegraphics[width=0.75\textwidth]{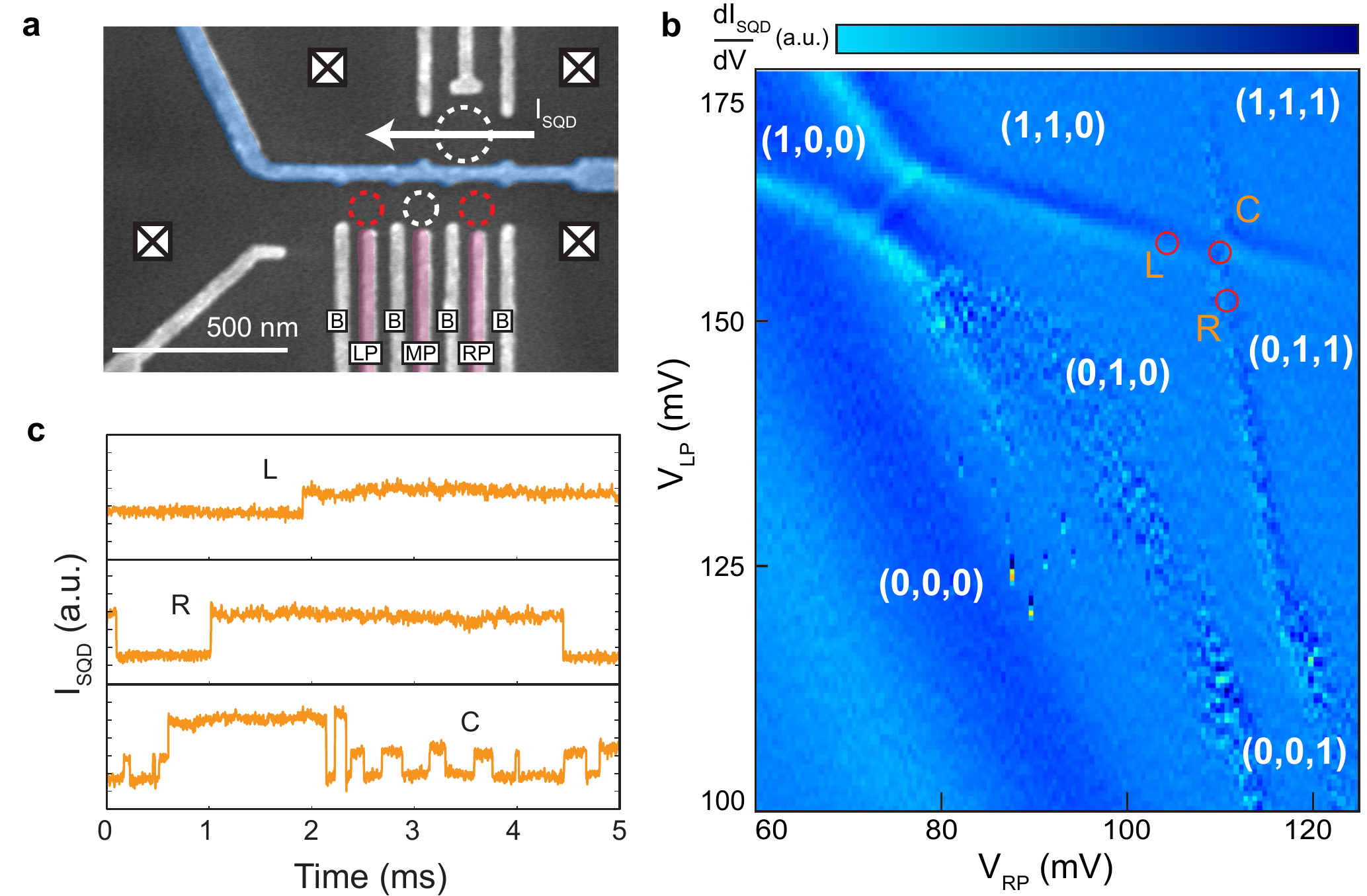}\\
\raggedright
\textbf{Figure 1: }\textbf{a}, SEM image of a sample identical to the one used for these measurements. Dotted circles indicate quantum dots, squares indicate Fermi reservoirs in the 2DEG, which are contacted through Ohmic contacts. Both the current through (white arrow) and the reflectance of the SQD is monitored and used to determine the occupancies of the triple quantum dot. \textbf{b}, Numerical derivative (along $V_{LP}$ axis) of the current through the SQD as a function of the voltages on gates LP and RP, mapping out a charge stability diagram of the triple dot in the few-electron regime. The (0,0,0)-(0,1,0) charging transition appears fragmented because of low tunneling rates from the reservoirs to the center dot. \textbf{c}, Real-time traces of the sensing dot reflectometry signal, taken at points L, R and C in Fig. 1b, as indicated. We use a 50 kHz low-pass filter (Avens Signal Equipment AP220) to filter the reflectometry signal in order to obtain sufficient signal-to-noise.
\vspace{0.4cm}
\label{fig:fig1}
\end{figure}

We operate the device in the few-electron regime: by sweeping the voltages on gates LP, MP and RP, the number of electrons on each of the dots of the triple dot array can be changed one by one. A time-averaged measurement of the differential DC-conductance through the SQD as a function of V$_{LP}$ and V$_{RP}$ maps out a cross-section of the three-dimensional charge stability diagram of the triple dot (Figure 1b). The occupancy is denoted here as $(n,m,p)$, corresponding to the number of electrons on the left, middle and right dot respectively. 

The strength of the four tunnel couplings can be tuned individually with the voltages on the B-gates. The tunnel rates between the outer dots and their respective leads are set to the 100 Hz range. The tunnel rates between neighbouring sites are tuned to be much higher. The upper two panels of Fig. 1c show real-time traces of the charge detector RF reflectometry signal taken at points R and L in the charge stability diagram. The first trace shows a single step, corresponding to the transfer of one electron from the left dot to the left reservoir, i.e. going from $(1,1,0)$ to $(0,1,0)$. In the second trace, three  single-electron tunnel events are seen, once from the right dot to the right reservoir and twice the other way, i.e. alternating between $(0,1,1)$ and $(0,1,0)$. The step size is larger than in the first trace, because of the closer proximity of the SQD to the right dot than to the left dot.

Remarkably, when we go to point C in the charge stability diagram, the real-time trace (lower panel) not only shows steps corresponding to the slow tunneling between outer dots and leads, but also exhibits smaller steps that occur at a rate which is an order of magnitude higher. Since point C is at the boundary of the $(1,1,0)$ or $(0,1,1)$ regions in the charge stability diagram, the fast steps appear to correspond to single electron transfers between the outer two dots. This is consistent with the step size as well as with the observation that the mean times that the measured conductance is high or low are equal for zero detuning between $(1,1,0)$ and $(0,1,1)$ (point C). Upon increasing or decreasing the detuning, these times quickly become unequal (see supporing online material).

This tunneling between the left and right dots is at first sight unexpected, since in these measurements the center dot levels are far detuned from resonance with the outer dot levels, excluding sequential tunneling via the center dot (there is no charging line of the center dot present nearby in the charge stability diagram). Here we argue that these tunneling events are transfers of single electrons between the outer dots, via \textit{virtual} occupation of the middle dot. This implies that electrons are transferred between the outer parts of the array, essentially without passing through the dot in between.

This tunneling between remote dots can be seen for different charge configurations of the triple dot array. Here we focus on transitions between $(1,1,0)$ and $(0,1,1)$. In this case, two virtual pathways exist for the transfer: either a single electron moves first virtually from the left to the middle dot and then from the middle to the right dot, or an electron moves first from middle to right and then another electron moves from left to middle (note that for other charge configurations the situation can be different, for instance for tunneling between $(1,0,0)$ and $(0,0,1)$, only the first pathway exists). As will be shown below, the existence of two virtual pathways makes the transfer rate depend non-monotonously on the detuning between the intermediate virtual states and the initial and final states. The dependence on the middle dot detuning is a key signature of this process.

\begin{figure}[htb]
\centering
\includegraphics[width=0.9\textwidth]{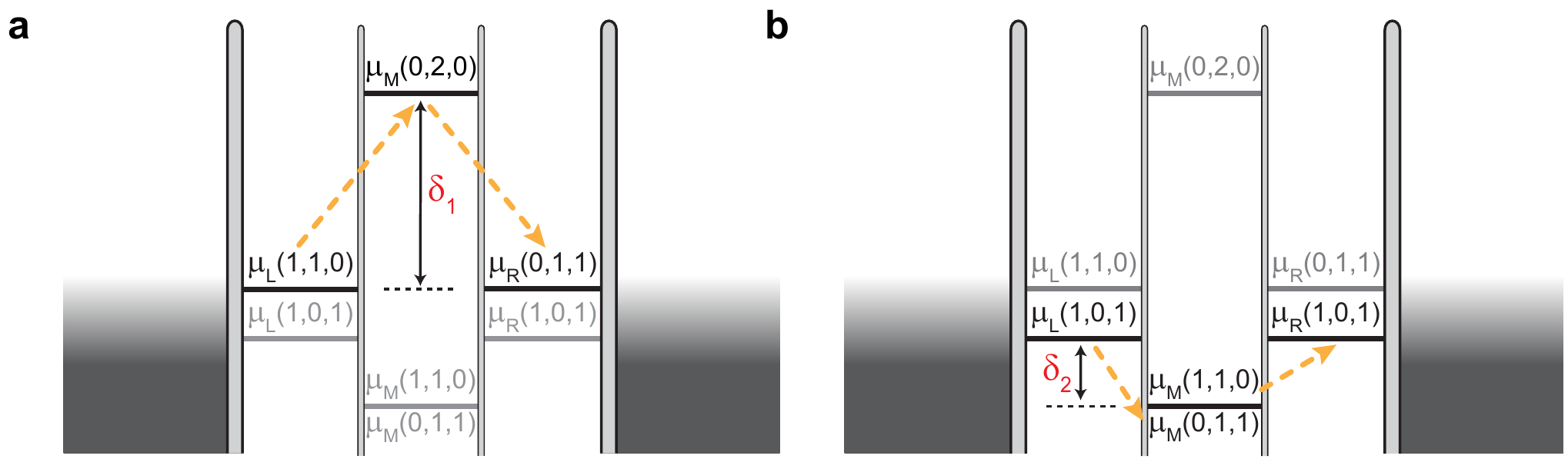}\\
\raggedright
\textbf{Figure 2: }\textbf{a, b}, Schematic representations of the cotunneling process in terms of the relevant electrochemical potentials in the linear dot array. The two panels illustrate the two possible pathways for cotunneling between $|110\rangle$ and $|011\rangle$, as explained in the main text.
\label{fig:fig2}
\end{figure}

\newpage
The charge transfer is depicted schematically in Figures 2a and 2b. Since it involves removing an electron from one dot and adding it to another dot, we need to compare electrochemical potentials for the transitions between initial, intermediate virtual, and final charge states. Only ground-state to ground-state transitions are considered. We denote the various electrochemical potentials as:

 \begin{centering}
	$\mu_L(n,m,p)=E(n,m,p)-E(n-1,m,p)$\\
	$\mu_M(n,m,p)=E(n,m,p)-E(n,m-1,p)$\\
	$\mu_R(n,m,p)=E(n,m,p)-E(n,m,p-1)$\\
\end{centering}
\noindent for the left, middle and right dot respectively. Here $E(n,m,p)$ stands for the ground-state energy of a given charge configuration $(n,m,p)$. For the transition back and forth between the states $|110\rangle$ and $|011\rangle$ to occur spontaneously, we require $\mu_L(1,1,0)=\mu_R(0,1,1)$ (we denote the lowest energy state with occupation $(n,m,p)$ as $|nmp\rangle$). Furthermore, in order to quantify the detuning between virtual states, and initial and final states, we introduce two parameters $\delta_1$ and $\delta_2$. For charge transfer via virtual occupation of $|020\rangle$, the relevant detuning parameter is $\delta_1=\mu_M(0,2,0)-\mu_L(1,1,0)$ (see Fig. 2a). For charge transfer via $|101\rangle$, the relevant detuning is $\delta_2=\mu_L(1,0,1)-\mu_M(1,1,0)$ (Fig. 2b). Note that $\delta_1$ and $\delta_2$ are related and cannot be changed independently, as $V_{MP}$ increases $\delta_1$ by the same amount it decreases $\delta_2$. The total tunnel rate $\Gamma$ is the sum of the tunnel rates via the two respective paths. It can be expressed as (see Supplementary Information):

\begin{equation}
\label{eq:rates}
\Gamma = \frac{2T_2}{\hbar}\left(\frac{t_{l1}^{2}t_{r1}^{2}}{\delta_{1}^{2}}+\frac{t_{l2}^{2}t_{r2}^{2}}{\delta_{2}^{2}}\right)
\end{equation}

Here $t_{l1}$, $t_{r1}$, $t_{l2}$ and $t_{r2}$ are the tunnel coupling elements between $|110\rangle$ and $|020\rangle$, $|020\rangle$ and $|011\rangle$, $\|101\rangle$ and $|011\rangle$, and $|110\rangle$ and $|101\rangle$, respectively. The charge dephasing time $T_2$ is assumed to be much smaller than $1/\Gamma$ ($T_2$ is typically about 1 ns~\cite{Hayashi03,Petta04} and therefore interference effects between the two pathways can be neglected). Equation 1 is valid as long as the four tunnel couplings and the detuning between $|110\rangle$ and $|011\rangle$ are small compared to $\delta_1$ and $\delta_2$.

We experimentally verify the non-monotonous dependence of $\Gamma$ on detuning $\delta_1$ (and hence $\delta_2$) by stepping the voltage on gate MP, $V_{MP}$, and measuring the rate of tunneling between $|110\rangle$ and $|011\rangle$. Figure 3a presents three traces, each for a different value of $V_{MP}$. For the top trace the value of $V_{MP}$ corresponds to small $\delta_1$ and large $\delta_2$, therefore the transfer proceeds mainly as depicted in Fig. 2a. For the middle trace, $V_{MP}$ is set such that both $\delta_1$ and $\delta_2$ are relatively large,  resulting in a reduced, but non-zero tunnel rate, in agreement with Eq.1. Finally, for the lower trace, $\delta_2$ is small and $\delta_1$ is large. In that case, tunneling proceeds mainly via the virtual process shown in Fig. 2b and the tunnel rate is higher again. 

\begin{figure}[htb]
\centering
\includegraphics[width=1\textwidth]{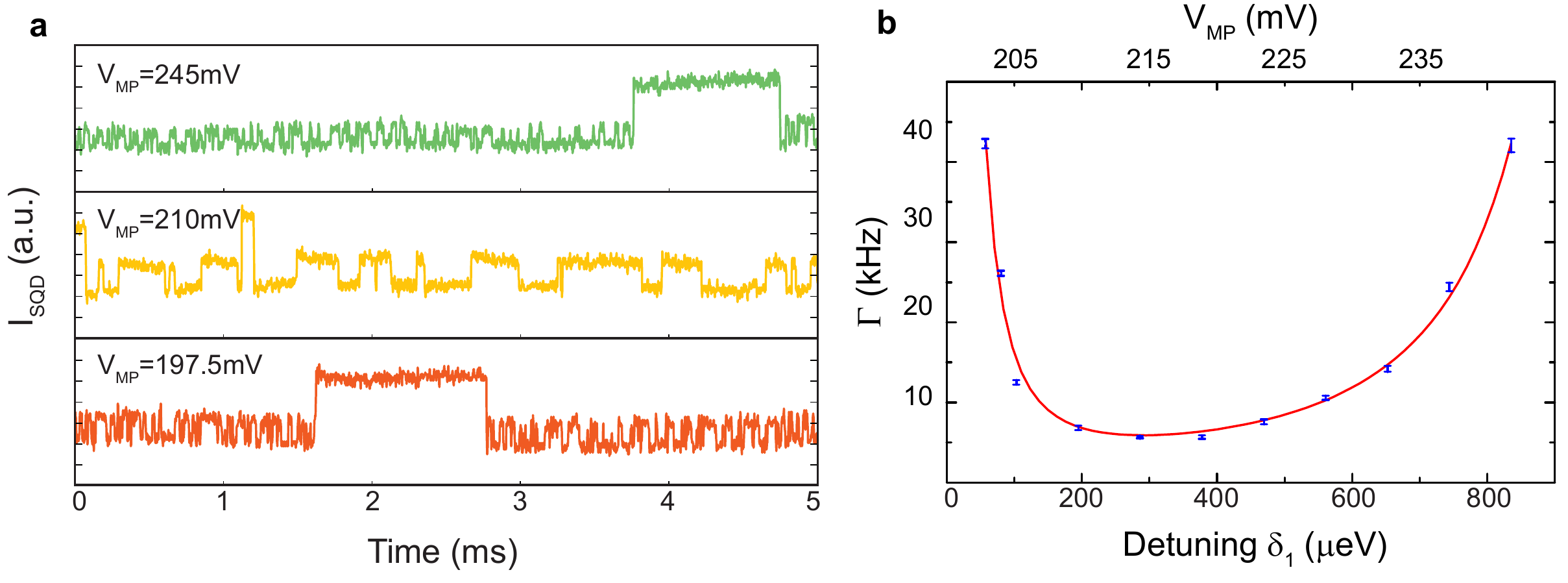}\\
\raggedright
\textbf{Figure 3: }\textbf{a}, Real-time traces of the SQD reflectometry signal, taken at zero detuning between the outer dot levels, for three values of $V_{MP}$, corresponding to three values of $\delta_1$ and $\delta_2$. \textbf{b}, Plot of the measured cotunneling rate $\Gamma$ versus detuning $\delta_1$. The non-monotonous dependence is a clear indication that the transfer proceeds via cotunneling. This is corroborated by the fact that the measured data points can be fitted well with the predicted expression for $\Gamma$ (red curve). To make the fit, we rewrite Eq.1 as $\Gamma = a/\delta_{1}^{2}+b/(c-d\delta_{1})^{2}$, with $a$, $b$, $c$ and $d$ positive constants. For the detuning axis, gate voltages are converted to energies using microwave-induced sidebands as an energy reference (see Supplementary Information). The error bars on the obtained values for $\Gamma$ include errors associated with the threshold analysis of the real-time traces (low-frequency noise modulates the baseline signal, so the precise value of the threshold slightly affects the statistics) and sampling errors due to the finite number of transfer events per trace\cite{Gustavsson09} (we sample over 100 ms traces). Note that the use of a low-pass filter results in an overall underestimation of $\Gamma$.
\label{fig:fig3}
\end{figure}

For a quantitative analysis, we extract the tunnel rate $\Gamma$ from real-time traces such as in Fig. 3a using the relation~\cite{Gustavsson09} $\Gamma^{-1}=f(1-f)\left(\langle\tau_{L}\rangle+\langle\tau_{R}\rangle\right)$, with $f$ the Fermi probability distribution. The times $\langle\tau_{L}\rangle$ and $\langle\tau_{R}\rangle$ are the average times between tunnel events spent in the left and right dot, respectively. We perform a threshold analysis of the real-time traces to obtain the distributions of $\tau_L$ and $\tau_R$. The value of $f$ can be established using the relation $f=\langle\tau_{L}\rangle/\left(\langle\tau_{L}\rangle+\langle\tau_{R}\rangle\right)$. Figure 3b shows measured values of $\Gamma$ thus determined, for different values of the detuning, parametrized by $\delta_1$. The non-monotonous dependence is striking and is fit well by Eq.1 (red curve), implying that the transfer indeed proceeds via virtual occupation of intermediate states. 

The hopping of electrons between the outer sites in the array indicates that an effective tunnel coupling is present between the left and right dot, which we call \textit{co}tunnel coupling. We can express the strength of this cotunnel coupling as (see Supplementary Information) \begin{equation}
\label{eq:coupling}
t_{co}=\frac{t_{l1}t_{r1}}{\delta_{1}}+\frac{t_{l2}t_{r2}}{\delta_{2}} \;.
\end{equation} 
We note that $t_{co}$ need not be small: for typical experimental values of the nearest-neighbour tunnel couplings of order 10 $\mu$eV and for detunings of $\sim$100 $\mu$eV, we obtain $t_{co} \sim$ 1 $\mu$eV. The cotunnel coupling enters the Hamiltonian in much the same way as the direct tunnel coupling between neighbouring sites and therefore many phenomena arising from direct tunnel coupling have their counterpart in cotunnel coupling between remote sites. For instance, upon application of microwave excitation to a gate, direct tunnel coupling can give rise to photon-assisted tunneling (PAT). In complete analogy we can expect a photon-assisted version of the cotunnelling process described above, which we term photon-assisted \textit{co}tunneling (PACT).

\begin{figure}[htb]
\centering
\includegraphics[width=0.75\textwidth]{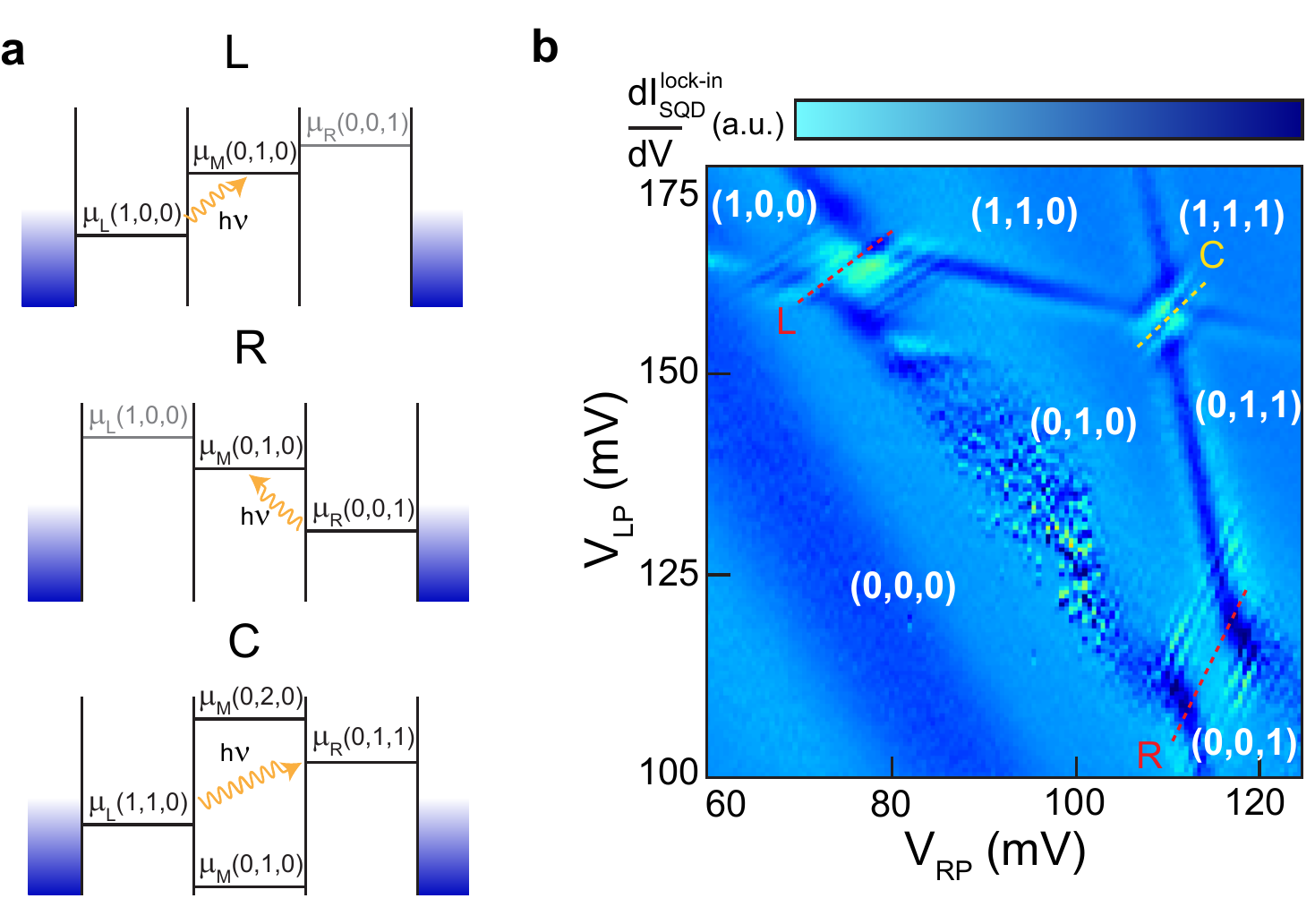}\\
\raggedright\textbf{Figure 4: }\textbf{a}, Schematic view of photon-assisted tunneling processes between different pairs of dots. Charges can be transferred from one dot to another when the detuning between the corresponding electrochemical potentials matches the photon energy. The left and middle panels correspond to PAT, the right panel corresponds to PACT. Note that similar resonances to the ones shown exist for negative detunings. \textbf{b}, Charge stability diagram in the same configuration as in Fig.1b, but now with microwave excitation (15 GHz) applied via a bias-tee to gate LP. The microwaves were chopped at the reference frequency of a lock-in amplifier and combined with a small amplitude modulation of the same reference frequency. The colorscale data is the numerical derivative (along $V_{LP}$ axis) of the SQD signal acquired via the lock-in amplifier. Multiple sidebands develop where PAT or PACT occurs.
\label{fig:fig4}
\end{figure}

Photon-assisted tunneling (PAT) is a well-described phenomenon in quantum dots and has been observed many times in single and tunnel coupled double quantum dots~\cite{Oosterkamp98,Wiel03}. Tunneling transitions of electrons between two detuned neighbouring dots can be made resonant by applying microwaves of a frequency matching the detuning, $\varepsilon_0 = nh\nu$ (see Figure 4a). Here $\varepsilon_0$ is the detuning, $\nu$ the microwave frequency and $n$ an integer, showing that also multiphoton resonances are possible. 

We apply microwave excitation at $\nu = 15$ GHz to gate LP. The microwaves are chopped at the reference frequency of a lock-in amplifier (see Supplementary Information). The excitation introduces a number of sidebands in the charge stability diagram. Two sets of sidebands are due to conventional PAT between neighbouring tunnel coupled quantum dots. They are seen in Figure 4b near the $|100\rangle$ to $|010\rangle$ transition (point L) and $|001\rangle$ to $|010\rangle$ transition (point R). The slope of these lines is such that $\varepsilon_0 = nh\nu$ is maintained.

Near the $|110\rangle$ to $|011\rangle$ transition (point C), a different set of resonances develop. Based on their slope and location, we identify these transitions to occur via PACT, where single electrons tunnel between the outer dots, now assisted by the microwave excitation. As expected for photon-assisted processes, these resonances appear at a detuning linearly dependent on the microwave frequency (see Supplementary Information).

In order to get a strong PACT response, the tunnel couplings between neighbouring dots are set to a much higher value than in the real-time experiment descibed before. We can extract the nearest-neighbour tunnel couplings from the spacing between the PAT resonances along the detuning axis as a function of frequency~\cite{Oosterkamp98,Wiel03} (see Supplementary Information) and find $8.1\pm0.4$ $\mu$eV for the tunnel coupling between right and center dot, and $12.3\pm0.3$ $\mu$eV between left and center dot. 

Importantly, photon-assisted cotunneling allows us to demonstrate the coherent dynamics driven by the cotunnel coupling, through the observation of Landau-Zener-St\"{u}ckelberg (LZS) interference~\cite{Shevchenko10}. Coherent quantum dynamics in the form of LZS interference has been observed in a wide range of two-level quantum systems, such as electronic states of atoms and molecules~\cite{Child74,Nikitin84} or superconducting devices~\cite{Oliver05} and spin states in double quantum dots~\cite{Petta10}. In LZS interferometry, a two-level system is swept through an anticrossing of its levels at such a rate that a superposition of its ground and excited states is reached (see Figure 5a). Between two passings through the anticrossing (at times $t_1$ and $t_2$), the two parts of the superposition acquire a relative phase due to their difference in energy $\varepsilon$, 
\begin{equation}
	\Delta\Theta_{12} = \frac{1}{\hbar}\int_{t_{1}}^{t_{2}}\varepsilon(t)dt \;.
\end{equation}
At the second passing through the anticrossing, the two paths in phase space will interfere. Destructive interference in the occupation probability of the excited state occurs for $\Delta\Theta_{12}=(2n+1)\pi$, where $n$ is an integer. 

\begin{figure}[htb]
\centering
\includegraphics[width=1\textwidth]{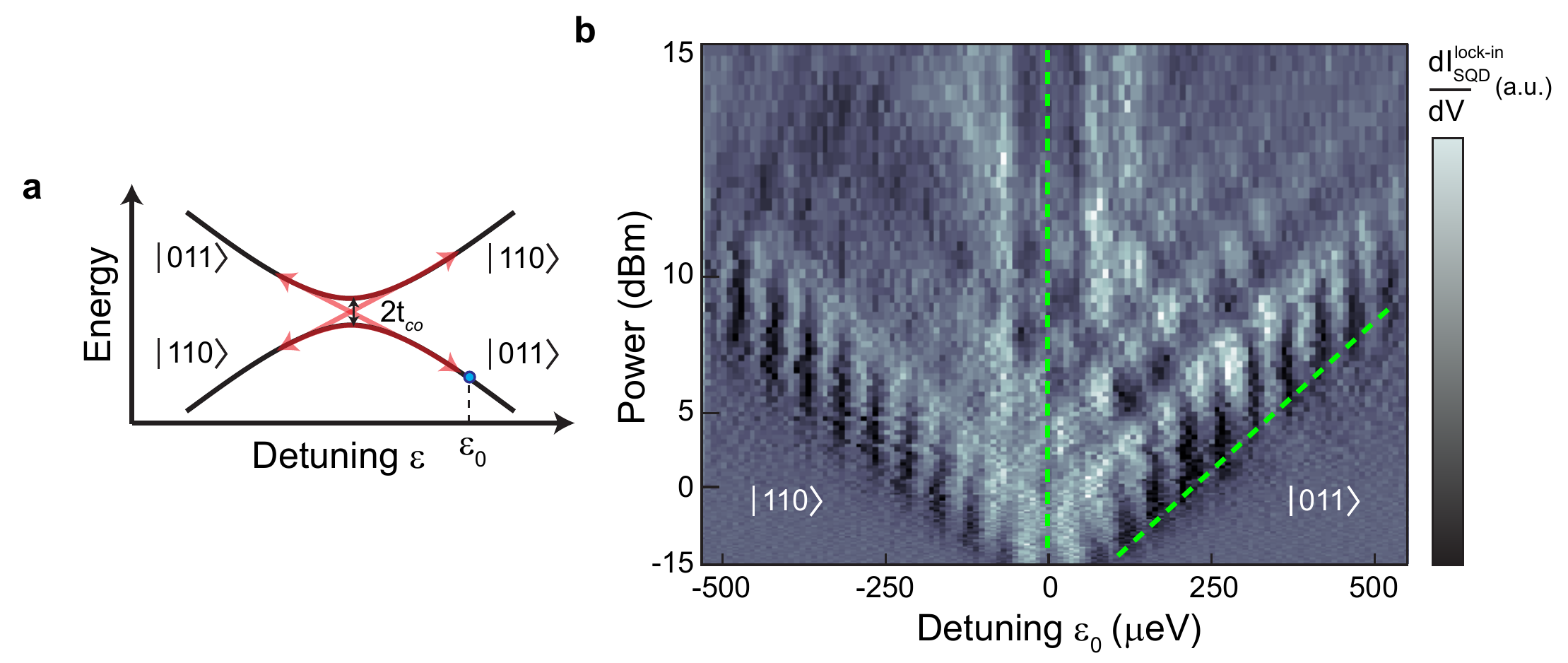}
\raggedright\textbf{Figure 5: }\textbf{a}, Schematic energy level diagram as function of detuning between $|110\rangle$ and $|011\rangle$, displaying an avoided crossing due to cotunnel coupling. The red arrows represent the response of the system to microwaves modulating the detuning. Multiple passings of the avoided crossing results in quantum interference of the two paths. \textbf{b}, Numerical derivative (along detuning axis) of the lock-in signal of $I_{SQD}$ (as in Figure 4b) as a function of detuning and microwave power. LZS-interference fringes are clearly visible along both axes.
\label{fig:fig5}
\end{figure}

In our experiment, the applied microwaves modulate the detuning $\varepsilon$ between $\mu_L(1,1,0)$ and $\mu_R(0,1,1)$. This takes the system back and forth through the anticrossing created by the cotunnel coupling (Fig. 5a). The amplitude of the microwaves is used to control the value of $\Delta\Theta_{12}$~\cite{Oliver05}. Figure 5b shows a measurement of the SQD signal as a function of detuning $\varepsilon_0$ versus microwave power. Contrast against the background indicates a finite population in the excited state (the ground state configuration for positive and negative detuning is as indicated in the figure). We see that for both positive and negative detunings, the excited state population exhibits interference fringes when sweeping the power (moving along the vertical dashed line). The oscillations indicate that a coherent superposition of $|110\rangle$ and $|011\rangle$ is created and maintained between subsequent passings through the anticrossing. 

Since the microwaves drive the system through the anticrossing periodically, also multiple crossings need to be considered. Consecutive cycles interfere constructively when the total phase difference accumulated over one complete microwave period $\Delta\Theta_{tot}$ equals $2\pi n$. As $\Delta\Theta_{tot}=\frac{1}{\hbar}\int \varepsilon(t)dt=2\pi\varepsilon_0/h \nu$, this causes peaks to appear for detunings $\varepsilon_{0,n} = n h \nu$. This can be clearly seen in Figure 5b, where along the horizontal direction ten fringes can be discerned.

In summary, we have demonstrated an effective coherent cotunnel coupling between the outer dots in a triple quantum dot, which is mediated by virtual occupation of levels on the dot in between. The coupling strength can be controlled through the detuning between the relevant middle and outer dot levels and agrees well with theoretical predictions. 

The long-range tunnel coupling may be used as well for realizing spin exchange gates at a distance in one step. When the intermediate site is itself occupied by an electron, its spin affects the strength of the cotunnel coupling due to the Pauli exclusion principle, introducing correlations between the middle spin and the outer spins (i.e. it realizes a three-qubit gate). First schemes for avoiding these correlations have been worked out, enabling direct long-range spin exchange also in this case \cite{privateloss13}. 
Furthermore, the cotunnel coupling can be extended to include multiple intermediate sites. Long-range coupling thus provides a new approach for operating quantum circuits based on quantum dot qubits, which eases the requirements for fault-tolerance. The cotunnel coupling observed here also gives access to a new range of phenomema with interacting spins, such as superexchange~\cite{Recher01,Mattis06} and frustration~\cite{Kim10}, which can serve as a starting point for quantum simulations.

\section*{Methods}
The experiment was performed on a GaAs/Al$_{0.25}$Ga$_{0.75}$As heterostructure grown by molecular beam epitaxy, with a 85 nm deep two-dimensional electron gas with electron density of $2.0\times10^{-11}$ cm$^{-2}$ and mobility of $5.6\times10^{-6}$ cm$^{2}$V$^{-1}$s$^{-1}$ at 4K. The metallic (Ti-Au) surface gates were fabricated using electron beam lithography. The device was cooled inside an Oxford AST Sorb dilution refrigerator to a base temperature of $\sim$55mK. In order to reduce charge noise~\cite{Pioroladriere05}, the sample was cooled down while applying a positive voltage bias on all gates, ranging between 200 and 350 mV. The magnetic field, as well as the bias across the linear triple quantum dot were set to zero throughout the experiment.  Gates LP and RP were connected to homebuilt bias-tees, enabling application of DC as well as high-frequency voltage bias to these gates. RF reflectometry of the SQD was performed using an LC circuit matching a carrier wave of frequency 193.35MHz. The power of the carrier wave arriving at the sample was about -84 dBm. The reflected signal was amplified using a cryogenic Quinstar QCA-U-219-33H amplifier and subsequently demodulated using homebuilt electronics. A Stanford Research Systems SR830 lock-in amplifier was utilized in some of the measurements. In these measurements, a square wave modulation of amplitude 2mV, before 16 dB attenuation, was applied to gate LP at the lock-in reference frequency of 3412Hz. For the microwave measurements, this square wave was combined with the microwaves, chopped at the same reference frequency. The microwaves were generated by an Agilent E8267D microwave source.

\newpage
\begin{centering}
{\Large Supplementary Information for} \\ \vspace{0.2cm}
{\Large \textbf{Long-range coherent coupling in a quantum dot array}}\\
\vspace{0.4cm}
\end{centering}

\section{Additional charge stability diagrams}
\begin{figure}[htb]
\centering
\includegraphics[width=1\textwidth]{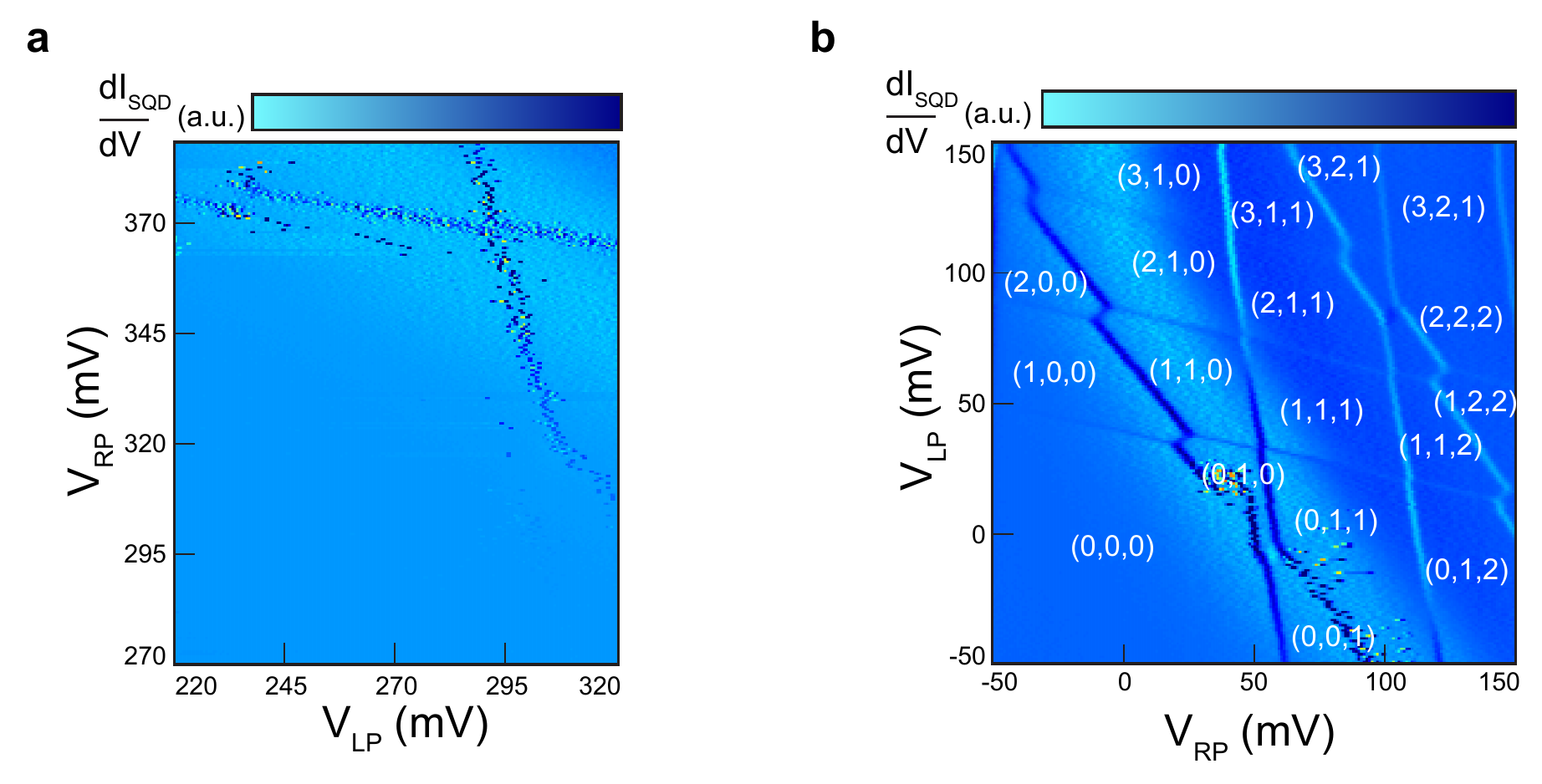}\\
\raggedright\textbf{Figure S1: }\textbf{a}, Numerical derivative (along horizontal axis) of the conductance of the SQD as a function of $V_{LP}$ and $V_{RP}$, measured in the opaque barrier regime of Fig.1c. \textbf{b}, Larger range measurement, displaying the numerical derivative (along horizontal axis) of the SQD conductance as a function of $V_{LP}$ and $V_{RP}$, this time in the more open barrier regime of Figs.1b and 4b.
\label{fig:sfig1}
\end{figure}

The real-time traces of Figure 1c and Figure 3a were taken in a different configuration of gate voltages than where the charge stability diagrams of Figure 1b and the PACT data of Figure 4b and Figure 5b were taken. Figure S1.a shows a charge stability diagram in this first regime, which was tuned such that all barriers (between dots and between dots and reservoirs) were quite opaque. Note that charging lines of the middle dot are not very visible, since it is charged at a very low rate due to the presence of multiple high barriers between this dot and the reservoirs. Figure S1.b shows a charge stability diagram in the more open barrier regime of Fig.4b, this time for a larger range of gate voltages than shown in the main text.

\section{Real-time traces for different detunings between outer dots}
\begin{figure}[htb]
\centering
\includegraphics[width=0.8\textwidth]{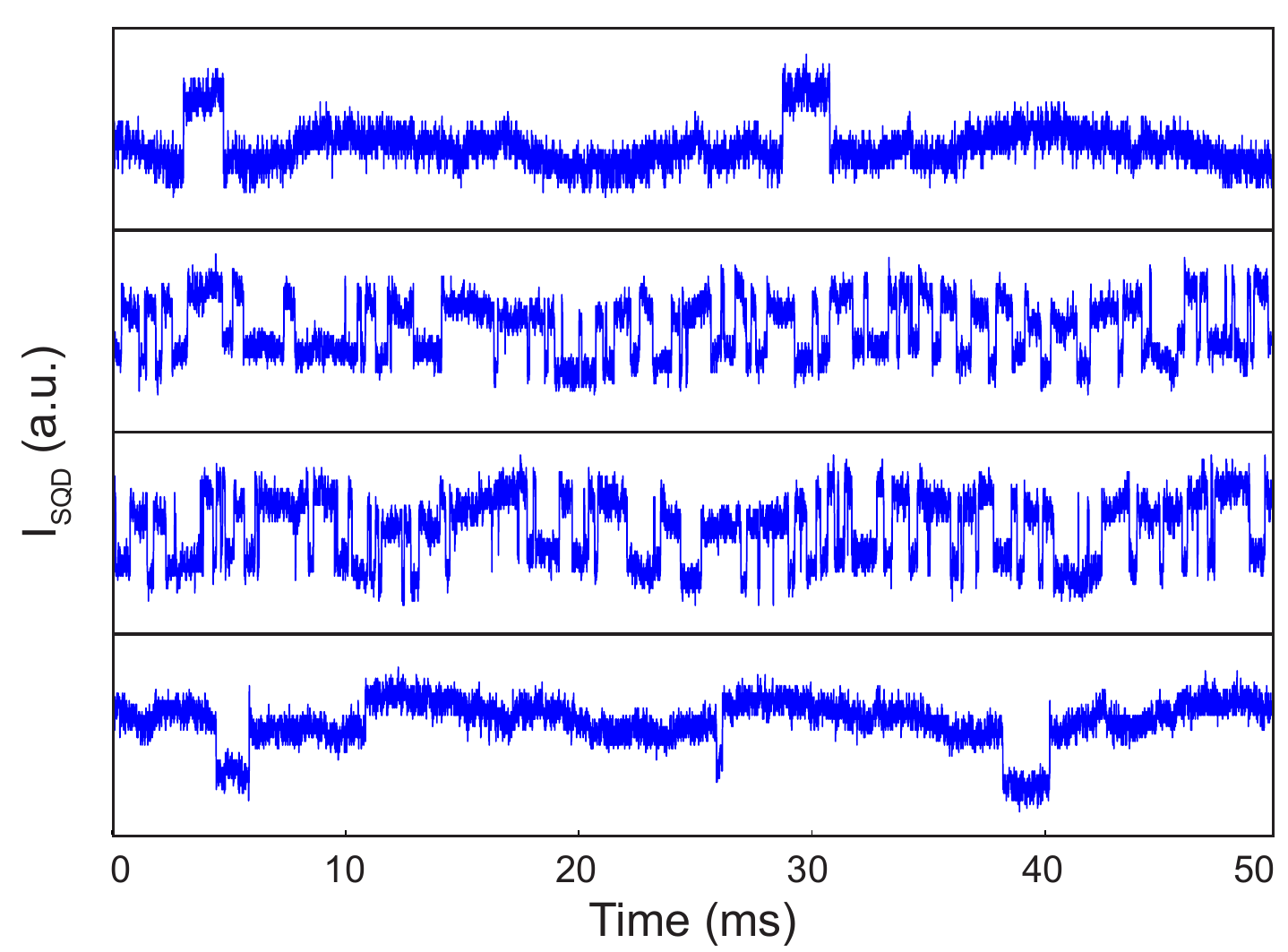}\\
\raggedright\textbf{Figure S1: }Real-time traces of the SQD signal for different detunings between $(1,1,0)$ and $(0,1,1)$, taken for $\delta_1\sim$300$\mu$eV. From top to bottom the detunings are: 49$\mu$eV, 12$\mu$eV, -24$\mu$eV and -61$\mu$eV.
\label{fig:sfig2}
\end{figure}

Supporting evidence that the tunneling events shown in Fig.1c are between the outer dots is given by the traces of Figure S2. The different traces are taken for different values of the detuning $\varepsilon$ between $(1,1,0)$ and $(0,1,1)$. From top to bottom, the detuning is increased from negative to positive values. In the upper trace, mainly $(0,1,1)$ is occupied. In the center traces, charges go back and forth most rapidly between the outer dots and in the lower trace, mostly $(1,1,0)$ is occupied.

\section{Cotunneling: effective tunnel coupling}

The Hamiltonian describing the experiments can be expressed in the basis $|\psi_1^{(0)}\rangle=|110\rangle, |\psi_2^{(0)}\rangle=|011\rangle, |\psi_3^{(0)}\rangle=|020\rangle, |\psi_4^{(0)}\rangle=|101\rangle$ as:
\begin{equation}
H=\begin{pmatrix}
    -\varepsilon/2 & 0 & t_{l1} & t_{l2} \\
    0 & \varepsilon/2 & t_{r1} & t_{r2} \\
    t_{l1} & t_{r1} & \delta_1 & 0 \\
    t_{l2} & t_{r2} & 0 & \delta_2
  \end{pmatrix}
\end{equation}

We perform a unitary transformation of this Hamiltonian, to express it in the eigenbasis of its first-order perturbation: $H'=U^\dagger H U$, where, as $\varepsilon<<\delta_{1,2}$:
\begin{equation}
U =\begin{pmatrix}
    1 & 0 & t_{l1}/\delta_1 & t_{l2}/\delta_2 \\
    0 & 1 & t_{r1}/\delta_1 & t_{r2}/\delta_2 \\
    -t_{l1}/\delta_1 & -t_{r1}/\delta_1 & 1 & 0 \\
    -t_{l2}/\delta_2 & -t_{r2}/\delta_2 & 0 & 1 
  \end{pmatrix} 
 \end{equation}
 
In this new basis, it is sufficient to only consider processes between the two first states:
 \begin{align}
|\psi_1^{(1)}\rangle&=|110\rangle-\frac{t_{l1}}{\delta_1}|020\rangle-\frac{t_{l2}}{\delta_2}|101\rangle \\
|\psi_2^{(1)}\rangle&=|011\rangle-\frac{t_{r1}}{\delta_1}|020\rangle-\frac{t_{r2}}{\delta_2}|101\rangle
\end{align}
Processes involving mixing between $\psi_1^{(1)}$ or $\psi_2^{(1)}$ and the other states are of third order in $t_i/\delta_j$ and can therefore be neglected. We can then reduce the Hamiltonian to: 
\begin{equation}  
H =\begin{pmatrix}
    -\frac{\varepsilon}{2}-\frac{t_{l1}^2}{\delta_1}-\frac{t_{l2}^2}{\delta_2} & \frac{t_{l1}t_{r1}}{\delta_1}+\frac{t_{l2}t_{r2}}{\delta_2} \\
    \frac{t_{l1}t_{r1}}{\delta_1}+\frac{t_{l2}t_{r2}}{\delta_2} & \frac{\varepsilon}{2}-\frac{t_{r1}^2}{\delta_1}-\frac{t_{r2}^2}{\delta_2}
    \end{pmatrix}
    =
    \begin{pmatrix}
    -\varepsilon'/2 & t_{co} \\
    t_{co} & \varepsilon'/2
    \end{pmatrix}
\end{equation}

This Hamiltonian then simply expresses an effective tunnel coupling $t_{co}=\frac{t_{l1}t_{r1}}{\delta_1}+\frac{t_{l2}t_{r2}}{\delta_2}$ between states whose detuning has been renormalized. In the presence of microwaves, we can write the detuning as $\varepsilon\rightarrow \varepsilon_0+Aexp(i\omega t)$, and the Hamiltonian becomes: 
\begin{equation}  
H    =
    \begin{pmatrix}
    -\varepsilon'_0/2-Aexp(i\omega t) & t_{co} \\
    t_{co} & \varepsilon'_0/2+Aexp(i\omega t)
    \end{pmatrix}
\end{equation}
which is exactly the Landau-Zener-St$\ddot{\mathrm{u}}$ckelberg Hamiltonian, so all the physics of the LZS-interference can be directly applied to describe photon-assisted cotunneling in the triple dot array.

\section{Lower bound estimation for charge $T_2$}
The width $w$ of the PACT resonances can be used to establish a lower bound on the charge dephasing time $T_2$, according to the relation~\cite{Shevchenko10}:
\begin{equation}
\label{eq:T2}
 w^2=\frac{t_{co}^2T_1}{T_2}+\frac{1}{T_{2}^{2}}>\frac{1}{T_{2}^{2}}
\end{equation}
With the measured value $w=4.47$ GHz, we find a lower boundary for $T_2$ of 224 ps. From PAT measurements between the left and middle and right and middle dots we extract lower bounds for $T_2$ of 394 ps and 338 ps respectively.

\section{Calculation of the real-time transition rate}

The relatively slow charge transition process in Figure 3 can be understood by means of a density matrix description. The evolution of the system is described by the quantum Liouville equation, $\frac{d\rho}{dt}=-i/\hbar[\rho,H]$, where H is the Hamiltonian and $\rho$ the density matrix. Given an effective tunnel coupling $t_{co}$ between two resonant states, the Hamiltonian is:
\begin{equation*}
H=\left(\begin{array}{cc}
  0 & t_{co} \\
  t_{co} & 0
\end{array}\right)
\end{equation*}

Adding phenomenologically decoherence occuring on a timescale $T_2$, we get the system of equations:
\begin{align*}
\frac{d}{dt}\rho_{11} &=-\frac{it_{co}}{\hbar}(\rho_{21}-\rho_{12})\\
\frac{d}{dt}\rho_{22} &=+\frac{it_{co}}{\hbar}(\rho_{21}-\rho_{12})\\
\frac{d}{dt}\rho_{12} &=-\frac{\rho_{12}}{T_2}-\frac{it_{co}}{\hbar}(\rho_{22}-\rho_{11})\\
\frac{d}{dt}\rho_{21} &=-\frac{\rho_{21}}{T_2}+\frac{it_{co}}{\hbar}(\rho_{22}-\rho_{11})
\end{align*}

We do not include relaxation channels, since processes inducing charge transitions other than the cotunneling are much slower than both cotunneling and decoherence.
Introducing $N=\rho_{22}-\rho_{11}$ and $P=\rho_{21}-\rho_{12}$, the equations reduce to
\begin{align*}
\frac{d}{dt}N  =&\frac{2it_{co}}{\hbar}P\\
\frac{d}{dt}P  =&\frac{P}{T_2}+\frac{2it_{co}}{\hbar}N
\end{align*}

We are interested in the rate with which charge moves back and forth between the outer dots. In our system, the initial conditions describe a pure state, where the electron is in the left dot: $N(0)=-1$, $P(0)=0$. The rate at which charge tunnels to the right dot is $d\rho_{22}/dt=\frac{1}{2}dN/dt$, where
\begin{equation*}
\frac{dN}{dt}=\frac{t_{co}^2T_2}{\hbar^2\sqrt{1-16t_{co}^2T_2^2/\hbar^2}}e^{-t/2T}\sinh\left(\frac{\sqrt{1-16t_{co}^2T_2^2/\hbar^2}}{2T_2}\right)
\end{equation*}
For a coherent process ($t_{co}>>1/T_2$) we obtain that a charge oscillates between the two dots at a rate linearly dependent on the coupling $t_{co}$:
\begin{equation*}
\frac{d\rho_{22}}{dt}=t_{co} e^{-t/2T_2}\sin\left(2\frac{t_{co} t}{\hbar}\right)
\end{equation*}

In the time traces of Fig. 1c and 3a, the barriers are tuned so that tunneling is extremely slow ($t_{co}<<1/T_2$), and the transition rate becomes
\begin{equation*}
\frac{d\rho_{22}}{dt}=2t_{co}^2T_2 (1-e^{-t/T_2})
\end{equation*}
As the timescale of the measurement is much longer than the decoherence time, the charge transition rate is therefore given by:
\begin{equation*}
\frac{1}{\tau}=2t_{co}^2T_2
\end{equation*}

In the experiments described in the manuscript, two paths ($|110\rangle\rightarrow|101\rangle\rightarrow|011\rangle$ and $|110\rangle\rightarrow|020\rangle\rightarrow|011\rangle$) contribute to the coupling $t_{co}$. For the slow, real time measurements of the tunneling rate, these two contributions add up incoherently and we have
\begin{equation*}
\frac{1}{\tau}=2t_{co}^2T_2=2T_2\left(\frac{t_{1}^4}{\delta_1^2}+\frac{t_{2}^4}{\delta_2^2}\right)
\end{equation*}
where $t_1=\sqrt{t_{l1}t_{r1}}$ ($t_2=\sqrt{t_{l2}t_{r2}}$) describes the coupling through $|101\rangle$ ($|020\rangle$), and $\delta_1$ and $\delta_2$ are the respective detunings.

\section{Calibration of the detuning between middle and outer dot levels in Fig. 3b}
\begin{figure}[htb]
\centering
\includegraphics[width=0.4\textwidth]{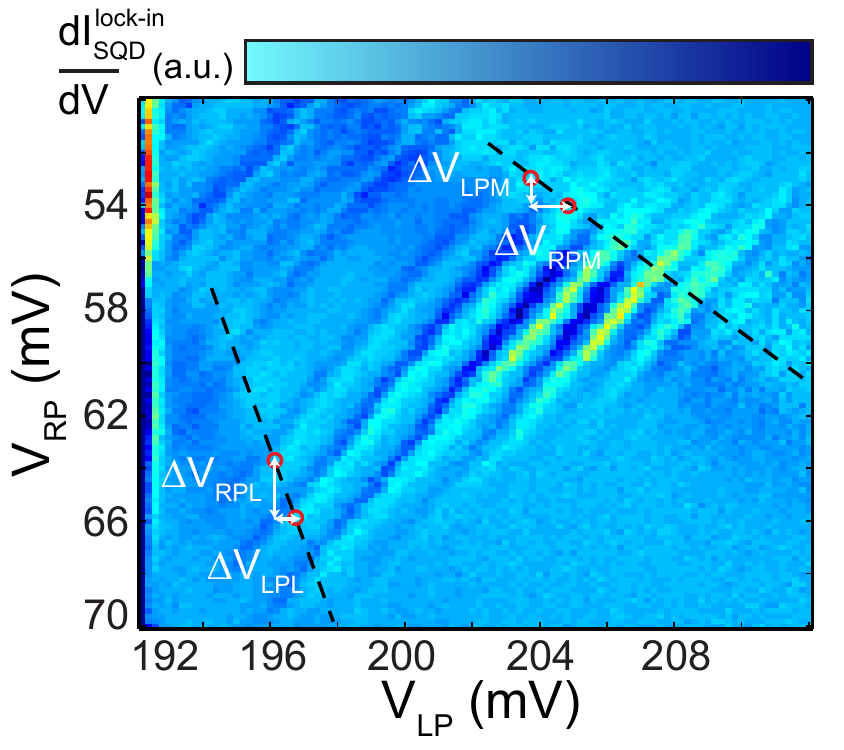}\\
\raggedright\textbf{Figure S3: }\textbf{a}, Numerical derivative (along horizontal axis) of the SQD lock-in signal (as in Fig.4b) as a function of $V_{LP}$ and $V_{RP}$, while microwaves of frequency 15 GHz are applied. The visible lines are the PAT resonances between the left and middle dot near point L in Fig.3b.
\label{fig:sfig3}
\end{figure}
We wish to determine the detunings $\delta_1$ and $\delta_2$ between the intermediate and initial/final electrochemical potentials for any set of gate voltages. To do so, we first convert gate voltages to energies via a set of conversion factors $\alpha_i^j$, where $i$ stands for LP, MP or RP and $j$ for LQD, MQD or RQD (left, middle and right quantum dot). Here $\alpha_i^j$ expresses by how much a change in the voltage applied to gate $i$ shifts the electrochemical potential of dot $j$. The values of the $\alpha_i^j$'s can be established using PAT measurements, using the known energy of the microwaves as a reference. Figure S3.a shows a measured charge stability diagram near the $(1,0,0)$-$(0,1,0)$ transition, with microwaves applied. There are multiple PAT sidebands visible. The distance between subsequent PAT resonances is set by the energy of the microwaves, $h\nu$. The slopes of the charging lines of each dot relate the relative influences of each gate on the electrochemical potentials on each dot: $S_{LQD}= \Delta V_{RPL}/\Delta V_{LPL} = -\alpha_{LP}^{LQD}/\alpha_{RP}^{LQD}$ and $S_{MQD}= \Delta V_{RPM}/\Delta V_{LPM} = -\alpha_{LP}^{MQD}/\alpha_{RP}^{MQD}$. Along the charging line of the left dot, the electrochemical potential of the involved state on that dot does not change: $d\mu_L(1,0,0)=0$. The distance between subsequent PAT resonances along that charging line then is related to a shift in electrochemical potential of the state of the other dot: $d\mu_M(0,1,0)=-\alpha_{LP}^{MQD}dV_{LPL}-\alpha_{RP}^{MQD}dV_{RPL}$. Using the expressions relating $\alpha_i^j$'s to slopes, we can derive: $\alpha_{LP}^{LQD}=-\frac{h\nu}{\Delta V_{LPM}}\frac{S_{LQD}}{S_{LQD}-S_{MQD}}$. Similar methods apply for the determination of the other conversion factors.
These conversion factors are used for determining the detuning axis of Fig.3b. The point in gate space where $\mu_L(1,1,0)=\mu_M(0,2,0)$ serves as a reference point, for which we define $\delta_1$ to be zero. When gate voltage $i$ is changed, the new value of $\delta_1$ is given by the gate voltage change multiplied by $\alpha_i^L - \alpha_i^M$. A similar reasoning applies to $\delta_2$.
 
\section{Frequency dependence of PAT and PACT}
\begin{figure}[htb]
\centering
\includegraphics[width=1\textwidth]{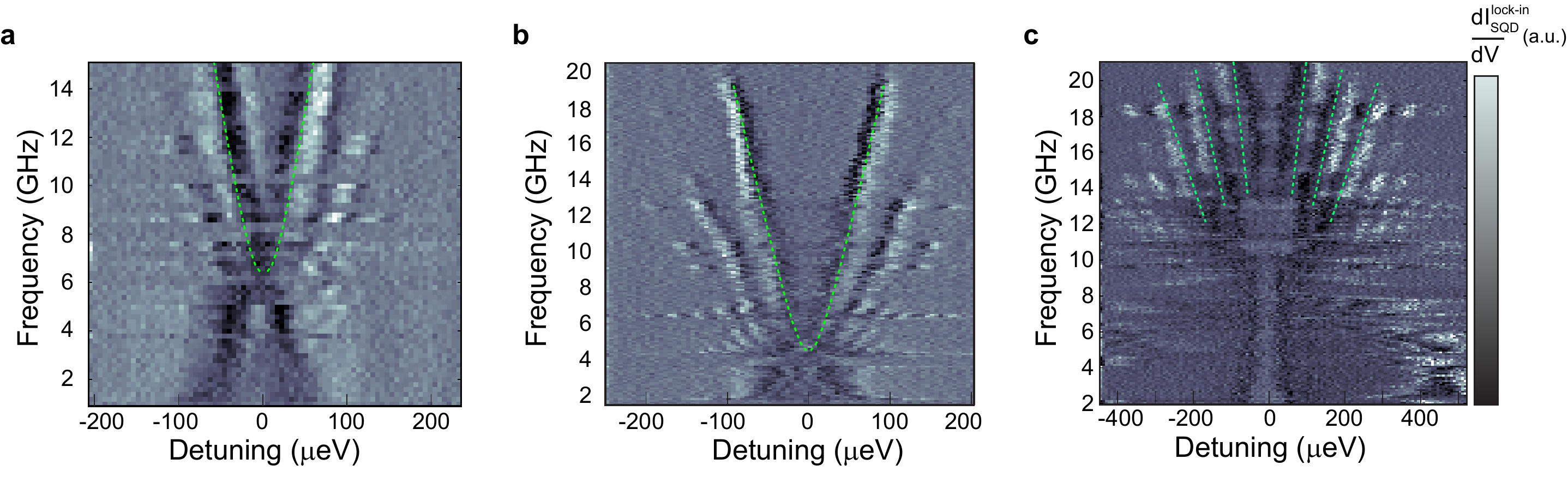}\\
\raggedright\textbf{Figure S4: }Numerical derivative (along horizontal axis) of the SQD lock-in signal (as in Fig.4b) as a function of frequency versus the detuning between: \textbf{a} $(1,0,0)$ and $(0,1,0)$, \textbf{b} $(0,0,1)$ and $(0,1,0)$, and \textbf{c} $(1,1,0)$ and $(0,1,1)$.
\label{fig:sfig4}
\end{figure}
From the frequency dependence of PAT between LQD and MQD (Figure S4.a), and between RQD and MQD (Figure S4.b), values for the tunnel couplings between these pairs of dots can be established, as in Oosterkamp et al~\cite{Oosterkamp98}. We find a tunnel coupling strength of $12.3 \mu eV$ between LQD and MQD and of $8.1 \mu eV$ between RQD and MQD, by fitting using $\Delta E = 2\sqrt{(h\nu)^2-(2t_c)^2)}$. As expected, for high driving frequencies, the frequency of the PACT resonances (Figure S4.c) shows a linear dependence on detuning, for each of the visible multiphoton resonances.

\end{document}